\documentclass[pdftex,twocolumn,epjc3]{svjour3}

\RequirePackage[T1]{fontenc}
\usepackage{amsmath}
\RequirePackage{graphicx}
\RequirePackage{mathptmx}      % use Times fonts if available on your TeX system
\RequirePackage{flushend}
\RequirePackage[numbers,sort&compress]{natbib}
\RequirePackage[colorlinks,citecolor=blue,urlcolor=blue,linkcolor=blue]{hyperref}

\usepackage{color}
\definecolor{darkgreen}{rgb}{0,0.5,0}
\definecolor{darkblue}{rgb}{0,0,0.7}
\definecolor{darkred}{rgb}{0.5,0,0.0}
\definecolor{darkorange}{rgb}{0.8,0.4,0.0}

\usepackage{amsmath}
\usepackage{amssymb}
\usepackage{xspace}
\usepackage{subfig}
\usepackage{comment}
\journalname{Eur. Phys. J. C}
\DeclareMathOperator{\De}{d}
\newcommand{\de}{\De\!}
\newcommand{\as}{\alpha_{\text{S}}}

\newcommand{\cf}{C_{\text F}}
\newcommand{\ca}{C_{\text A}}

\newcommand{\kt}{k_t}

\newcommand{\muf}{\mu_{\text{F}}}

\newcommand{\muOf}{\mu_{0\text{F}}}
\newcommand{\mur}{\mu_{\text{R}}}

\newcommand{\muOr}{\mu_{0\text{R}}}

\usepackage{cprotect}

\begin{document}

\title{An improved description of charm fragmentation data}
\author{
Matteo Cacciari\thanksref{e1,addr1,addr3}
\and
Andrea Ghira\thanksref{e2,addr2}
\and
Simone Marzani\thanksref{e3,addr2}
\and
Giovanni Ridolfi\thanksref{e4,addr2}
}

\thankstext{e1}{e-mail: cacciari@lpthe.jussieu.fr}
\thankstext{e2}{e-mail: andrea.ghira@ge.infn.it}
\thankstext{e3}{e-mail: simone.marzani@ge.infn.it}
\thankstext{e4}{e-mail: giovanni.ridolfi@ge.infn.it}

\institute{Sorbonne Universit\'e, CNRS, Laboratoire de Physique Th\'eorique et Hautes \'Energies,
LPTHE, F-75005 Paris, France\label{addr1}
\and
Universit\'e Paris Cit\'e, LPTHE, F-75006 Paris, France\label{addr3}
\and
Dipartimento di Fisica, Universit\`a di Genova and INFN, Sezione di Genova, Via Dodecaneso 33, 16146, Italy\label{addr2}
}

\date{}

%\institute{teste}

\maketitle

\begin{abstract}
We consider the fragmentation of heavy quarks into heavy-flavoured hadrons, specifically the production of charmed mesons in $e^+e^-$ collisions, at different centre-of-mass energies. 
We focus our attention on the ratio of moments of the $D^{*+}$ energy spectrum measured by ALEPH and CLEO.
This ratio is believed to provide us with a direct test of perturbative QCD evolution because hadronisation effects should cancel between the numerator and denominator.  However, state-of-the-art calculations based on standard (final-state) collinear factorisation fail to describe the experimental data. We show that this discrepancy is considerably reduced if heavy-quark threshold effects are accounted for not only in DGLAP evolution, as it is usually done, but also in the resummed coefficient functions. 
\end{abstract}
%%%%%%%%%%%%%%%%%%%%%%%%%%%%%%%%%%%%%%%%%%%%%%%%%%%%%%%%%%%%%%%%%%%%%%%%%%
% Letters should not exceed 4 printed pages in the EPJ style format, and should contain no more than 4 figures and/or tables.
Studies of fragmentation functions of heavy quarks, namely charm ($c$) and bottom ($b$), into heavy-flavoured hadrons, e.g.\ $D$ and $B$ provide stringent tests of our understanding and modelling of strong interactions, in the presence of disparate energy scales. 
Similar to what happens for light-quark fragmentation, the evolution from the scale of the hard interaction down to the scale at which the hadronisation process turns quarks into observable hadrons can be described in perturbative Quantum ChromoDynamics (QCD) thanks to DGLAP equations, with time-like splitting functions. However, unlike the light-quark case, the masses of $c$ and $b$ are in the perturbative regime. Thus, for heavy flavours, the low-scale initial condition of the evolution can also be determined perturbatively.

Perturbative calculations are plagued by large logarithmic corrections of different kinematic origins. DGLAP evolution, together with the determination of the initial condition at the appropriate order, allows for the resummation of mass logarithms of collinear origin. The next-to-leading order (NLO) calculation was performed in~\cite{Mele:1990yq,Mele:1990cw} and then extended to NNLO in~\cite{Rijken:1996vr,Mitov:2006wy,Blumlein:2006rr,Melnikov:2004bm,Mitov:2004du}.
On the other hand, one also encounters logarithms of soft origin that need to be resummed too. This is particularly relevant for heavy-quark production because, as it was realised in the early days, fragmentation functions of heavy quarks peak at large hadron energy fractions~\cite{Suzuki:1977km,Bjorken:1977md,Kinoshita:1981af,Kinoshita:1985mh,Peterson:1982ak}. 
The resummation of soft logarithms was performed at leading logarithmic accuracy (LL) in~\cite{Mele:1990yq,Mele:1990cw}, extended to NLL in~\cite{Cacciari:2001cw} and it is currently known to NNLL~\cite{Fickinger:2016rfd,Ridolfi:2019bch,Maltoni:2022bpy,Czakon:2022pyz}. Central for the results presented in this letter, is the nontrivial interplay between the resummation of mass logarithms and soft ones~\cite{Aglietti:2007bp,Aglietti:2022rcm,Gaggero:2022hmv,Ghira:2023bxr}. 

To obtain phenomenological predictions that can be compared to experimental data, perturbative calculations must be supplemented with non-perturbative corrections that account for the hadronisation process. Several approaches exist in the literature, exploiting effective theories~\cite{Jaffe:1993ie,Neubert:2007je,Fickinger:2016rfd}, renormalons~\cite{Gardi:2003ar,Gardi:2005yi}, and effective strong coupling~\cite{Aglietti:2006yf,Corcella:2007tg}.

Detailed phenomenological studies for both $b$ and $c$ fragmentations in electron-positron ($e^+e^-$) collisions have been performed at NLO+NLL~\cite{Cacciari:2005uk} and, more recently, at NNLO+ NNLL~\cite{Bonino:2023icn}. These studies constitute the starting point of this letter. In particular, we focus on charm fragmentation and show that by supplementing the standard NNLO+NNLL calculation of Ref.~\cite{Bonino:2023icn} with the consistent resummation of mass and soft logarithms developed in Ref.~\cite{Ghira:2023bxr}, we can alleviate a long-standing tension~\cite{Cacciari:2005uk,Bonino:2023icn} between the theoretical predictions and experimental data. 

In this study, we consider Mellin moments of the (normalised) energy spectrum for a heavy-flavoured hadron $H$:
\begin{equation}
    \Sigma(N,Q^2)=\frac{1}{\sigma}\int_0^1 dx\,  x^{N-1}\frac{d \sigma}{d x},
\end{equation}
where $x=E_H/E_\text{beam}=2 E_H/Q$~\footnote{Because $\frac{2 m_H}{Q}<x<1$, lower-energy experiments often consider the momentum fraction $x_p=\sqrt{\frac{x^2 Q^2-4m_H^2}{Q^2-4 m_H^2}}$, with $0< x_p <1$. The difference is a power correction $\frac{m_H^2}{Q^2}$, beyond the accuracy of the calculation presented in this Letter.}.
In the fragmentation function approach the mass of the heavy quark $m_h$ sets a perturbative scale and is retained only as the regulator of the collinear singularities, while power corrections $\frac{m_h^2}{Q^2}$ are neglected. Within this approximation, we can write Mellin moments of the spectrum in a factorised form
\begin{align}\label{eq:FF-factorisation-1}
\Sigma(N,Q^2)&=\sigma_h(N,Q^2) D^\text{NP}_{h\to H}(N)
\end{align}
with
\begin{align}\label{eq:FF-factorisation-2}
\sigma_h(N,Q^2)&= \hspace{-0.25 cm}\sum_{i=q,h,g} C_i\left(N,Q^2,\mu^2\right) D_{i\to h}\left(N,\mu^2,\mu_0^2\right) +\mathcal{O}\left(\frac{m_h^2}{Q^2} \right)
\end{align}
where $\sigma_h$ is perturbative cross section for the production of a heavy quark $h$ and the sum in Eq.~(\ref{eq:FF-factorisation-2}) runs over all massless partons, including the nominally heavy quark $h$.  
The first contribution is the perturbative coefficient function for the production of a massless parton $i$ and it is evaluated at the scale $\mu^2\simeq Q^2$.\footnote{We can actually introduce two kinds of scales: the renormalisation scale $\mur$ and the factorisation scale $\muf$. To keep the notation simple, we have set $\mur=\muf=\mu$, and $\muOr=\muOf=\mu_0$.}
The second contribution is the heavy-quark fragmentation function evolved from the low scale $\mu_0^2\simeq m_h^2$ to the hard scale $\mu^2\simeq Q^2$:
\begin{equation}\label{eq:DGLAPfull}
    D_{i\to h}\left(N,\mu^2\right)=\sum_j E_{ij}(N,\mu^2,\mu_0^2)D_{j\to h}(N,\mu_0^2,m_h^2).
\end{equation}
Note that the initial condition is defined in a scheme in which the heavy flavour $h$ is not active, while it does contribute to the evolution. It may happen that during the evolution a subsequent quark threshold is crossed. In such a case, the number of active flavours is increased by one unit and appropriate matching conditions are adopted~\cite{Cacciari:2005ry}.
Finally, $D^\text{NP}_{h\to H}(N)$ describes the non-perturbative correction that must be fitted to experimental data. 

An exceptionally interesting observable to study is the ratio of the above moments for the same process but at experiments performed at different centre-of-mass energies
\begin{equation}\label{eq:ratio_def}
    R=\frac{ \Sigma(N,Q_A^2)}{ \Sigma(N,Q_B^2)}.
\end{equation}
Specifically, we consider the $e^+e^-$ production of the charmed meson $D^{*+}$ as measured by the ALEPH experiment at LEP at the $Z$ boson peak, $Q_A=m_Z$~\cite{ALEPH:1999syy} and by the CLEO collaboration near the $\Upsilon(4S)$ resonance, $Q_B=m_{\Upsilon(4S)}$~\cite{CLEO:2004enr}. \footnote{Refs.~\cite{Cacciari:2005uk,Bonino:2023icn} also considered analogous data collected by BELLE~\cite{Belle:2005mtx}. However, these data have been withdrawn by the collaboration. Therefore, while we do not see any significant difference between BELLE and CLEO data, we will not consider them here.}

\begin{figure}
	%\centerings
\includegraphics[width=0.48\textwidth]{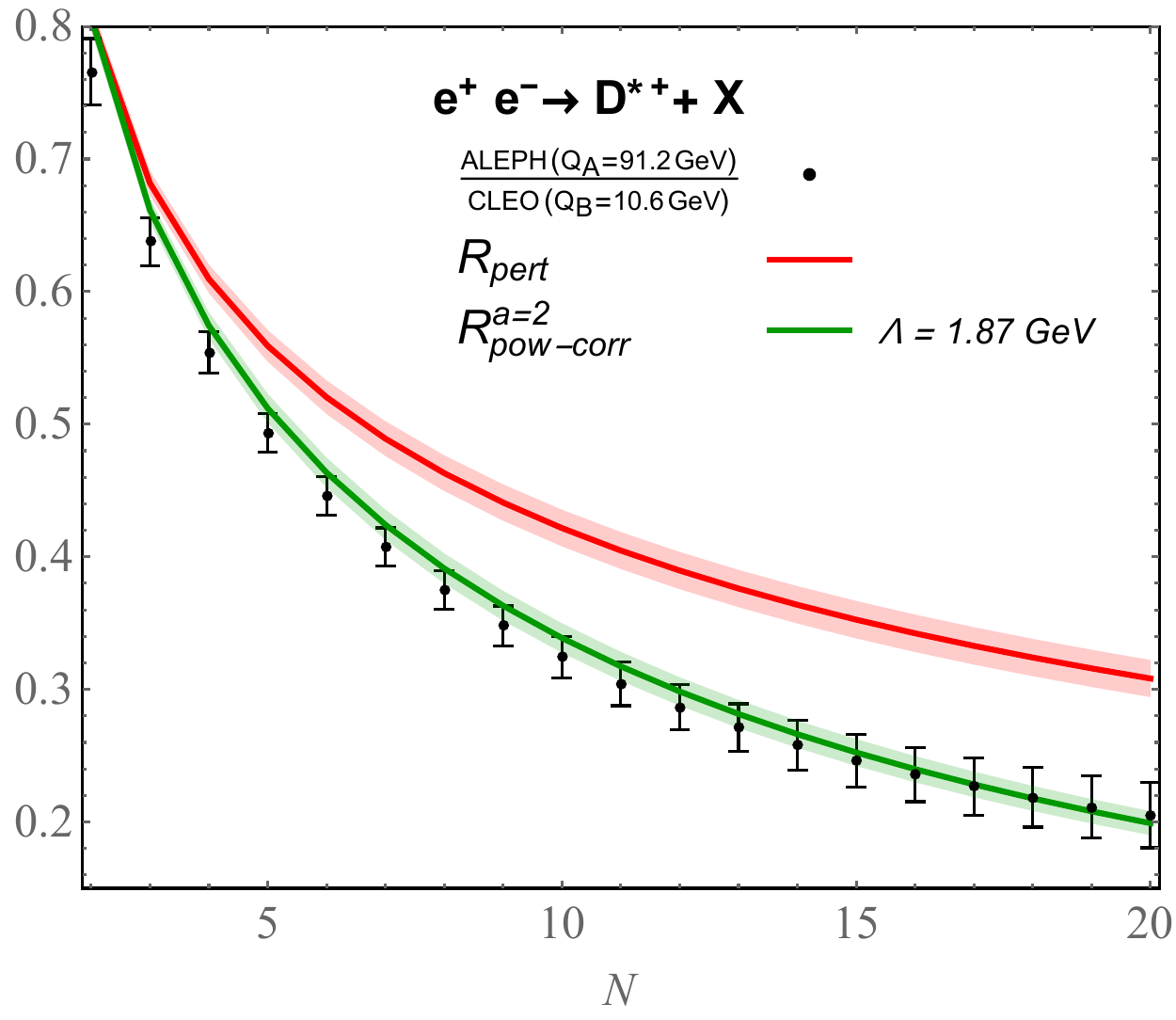}
\vspace{-0.5cm}
	\caption{Ratio of moments of $D^{*+}$ energy spectrum in $e^+e^-$ collisions at two different centre-of-mass energies: $Q_A=91.2$~GeV and $Q_B=10.6$~GeV. The experimental data points, in black, are obtained by taking the ratio of ALEPH and CLEO measurements. The NNLO+NNLL theoretical prediction $R_\text{pert}$ is obtained from Eq.~(\ref{eq:ratio_cntd}), with $m_c=1.5$~GeV, $m_b=4.75$~GeV, and it is shown in red. The green curve instead corresponds to the NNLO+NNLL together with the non-perturbative factor with quadratic power corrections, i.e. $ R_\text{pow-corr}^{a=2}$, Eq.~(\ref{eq:R_nonpert}) and fitted $\Lambda=1.87$~GeV.}
	\label{fig:bcs}
\end{figure}
We expect the contributions from low-scale physics, i.e.\ the initial condition and its non-perturbative correction, to largely cancel in the ratio of Eq.~(\ref{eq:ratio_def}). In particular, the non-perturbative correction cancels, while the perturbative initial condition does so if we restrict ourselves to the flavour non-singlet component, which constitutes the dominant contribution at large $N$~\cite{Cacciari:2005uk}:
\begin{align}\label{eq:ratio_cntd}
    R_\text{pert}&=\frac{\sum_{i=q,c,g} C_i\left(N,Q^2_A,\mu_A^2\right) D_{i\to c}\left(N,\mu_A^2,\mu_0^2\right) D^\text{NP}_{c\to D^{*+}}(N)}{\sum_{i=q,c,g} C_i\left(N,Q_B^2,\mu_B^2\right) D_{i\to c}\left(N,\mu_B^2,\mu_0^2\right) D^\text{NP}_{c\to D^{*+}}(N) }\nonumber\\
    &\simeq \frac{C_h\left(N,Q^2_A,\mu_A^2\right)}{C_h\left(N,Q^2_B,\mu_B^2\right)}E(N,\mu_A^2,\mu_B^2), \; \text{with}\; \mu_{A,B}^2\simeq Q_{A,B}^2.
\end{align}
Therefore, in the non-singlet approximation, the charm ratio should provide us with a direct test of perturbative evolution in QCD. 
%
%Note that the same mechanism also appears in light-quark fragmentation. However, this case is less relevant for phenomenology because the non-singlet component is not as dominant.

The comparison of the theoretical prediction for the charm ratio to the experimental data was first performed at NLO+ NLL in~\cite{Cacciari:2005uk} and repeated at the subsequent order in~\cite{Bonino:2023icn}.~\footnote{Mellin moments of the experimental data are corrected for QED initial-state radiation, according to the procedure described in~\cite{Cacciari:2005uk}.} 
The NNLO+NNLL results obtained with the code developed in \cite{Bonino:2023icn}, which also uses~\cite{Bertone:2015cwa} for DGLAP evolution, are shown in Fig.~\ref{fig:bcs}. The input parameters are as follows: $Q_A=91.2$~GeV, $Q_B=10.6$~GeV, $m_b=4.75$~GeV, $m_c=1.5$~GeV, $\as(m_Z^2)=0.118$. The uncertainty bands are obtained by reinstating separate $\mur,\muOr,\muf$, and $\muOf$ dependence and by varying them independently, as detailed in~\cite{Bonino:2023icn}. The only difference with respect to the original setup is the resummation of logarithms of $\bar N= N e^{\gamma_E}$ rather than of $N$. At this order, this has a tiny effect on the central value and leads to a small reduction of the theoretical uncertainty. 

The surprising result is a large discrepancy between theory and data. The discrepancy grows larger as $N$ increases. In this region, which corresponds to $x \to 1$ in momentum space, the ratio is dominated by the non-singlet and, therefore, the cancellation described in Eq.~(\ref{eq:ratio_cntd}) should hold. 
In~\cite{Cacciari:2005uk} it was speculated that the origin of such a discrepancy could be ascribed to non-perturbative power corrections to the coefficient functions, leading to
\begin{equation}\label{eq:R_nonpert}
    R_\text{pow-corr}^{a}=R_\text{pert}\times \frac{1+\frac{\Lambda^a}{Q_A^a}\mathcal{C}(N)}{1+\frac{\Lambda^a}{Q_B^a}\mathcal{C}(N)},
\end{equation}
where we expect $\mathcal{C}(1)=0$ and, from infrared renormalon calculations $a=2$~\cite{Dasgupta:1996ki,Beneke:1997sr}. If we make the same ansatz $\mathcal{C}(N)=N-1$ as~\cite{Cacciari:2005uk}, a one-parameter fit to the data returns the value $\Lambda=1.87$~GeV for the non-perturbative scale, slightly larger than expected. 
%the expected $\simeq 1$~GeV. 

We note that perturbative power corrections $\frac{m_c^2}{Q^2}$ are instead unlikely to explain the discrepancy because their coefficient cannot be linear in $N$ but can grow at most as a power of $\log N$, see e.g.~\cite{Nason:1999zj} in which a matched calculation of massive and massless predictions show that these corrections are of limited importance. However, as recently pointed out in~\cite{Ghira:2023bxr}, when considering resummed spectra there can be subtle perturbative effects to be considered. The main result of~\cite{Ghira:2023bxr}, when restricted to real values of the Mellin variable $N$, is an all-order expression defined piecewise that, in each region, consistently resume both logarithms of $N$ and logarithms of the heavy-quark mass, to NLL.
In particular, two distinct effects are accounted for. The first one is the so-called dead cone effect~\cite{Dokshitzer:1991fd,Dokshitzer:1995ev} for emissions collinear to the unmeasured (anti)quark. Following the arguments presented in~\cite{Ghira:2023bxr}, this contribution becomes relevant when 
\begin{equation}\label{eq:moments30}
N> \frac{m_{\Upsilon(4S)}^2 }{m_c^2}e^{-\gamma_E}\simeq 30.    
\end{equation}
This is outside the region considered in Fig.~\ref{fig:bcs} and therefore not relevant to this discussion. 

The second effect has to do with the treatment of heavy-quark thresholds in the resummed calculation of the perturbative coefficient function. When performing the all-order calculation for $C_h(N)$ in the soft limit, we have to perform integrations over the running coupling, evaluated at the characteristic scale of each splitting, namely the relative transverse momentum $\kt$. Thus, if we adopt the decoupling scheme, we have to use
\begin{align}\label{eq:rc-decoupling}
    \as(\kt^2)&= \as^{(5)}(\kt^2)\Theta(\kt^2-m_b^2) \nonumber \\&+\as^{(4)}(\kt^2)\Theta(m_b^2-\kt^2)\Theta(\kt^2-m_c^2)\\
     \nonumber &+\as^{(3)}(\kt^2)\Theta(m_c^2-\kt^2).
\end{align}

We point out a couple of differences with respect to the calculation detailed in~\cite{Ghira:2023bxr}.
First, we note that because we are interested in charm fragmentation at scales $(Q_A,Q_B)$ above the bottom mass, we have to generalise the original calculation that only considered a single threshold.
Second, rather than writing the result as the product of a measured jet function and a recoil one, as done in~\cite{Ghira:2023bxr}, we find it more convenient to express it as 
\begin{equation}\label{eq:GMR-start}
    \log C_\text{thr}(N,Q^2,\muf^2)=\Delta(N,Q^2,\muf^2)+\bar J(N,Q^2),
\end{equation}
with
\begin{align}\label{eq:DeltaJbar}
 \Delta(N,Q^2,\muf^2)&=\int^1_{\frac{1}{\bar{N}}} \frac{\de z}{z}\int^{\muf^2}_{z^2 Q^2} \frac{\de \kt^2}{\kt^2} A(\as(\kt^2))
    \nonumber\\
    \bar J(N,Q^2)&=-\int^1_{\frac{1}{\bar{N}}} \frac{\de z}{z}\Big[\int^{z Q^2}_{z^2 Q^2} \frac{\de \kt^2}{\kt^2} A(\as(\kt^2))\\&\quad \quad \quad+\frac{1}{2}B(\as(z Q^2))\Big]\Theta \left( \frac{Q^2}{m_c^2}-\bar N \right),\nonumber
\end{align}
where the dependence on the renormalisation scales is understood.
The functions $A,B$ have an expansion in powers of $\as$:
\begin{equation}\label{Resummation_functions}
 	\begin{split}
 		A(\as)=\sum_{k} \left(\frac{\as}{\pi}\right)^k A_{k},\quad	B(\as)=\sum_{k}\left(\frac{\as}{\pi}\right)^k B_{k}.
 	\end{split}
 \end{equation}
 To NLL we need $A_{1}=\cf$, $A_{2}^{(n)}=\frac{\cf \ca}{2} \left(\frac{67}{18}-\zeta_2\right)-\frac{5}{9} n$, $ B_{1}=-\frac{3}{2}\cf$, where $n$ is the number of active flavours.
All the integrals in Eqs.~(\ref{eq:DeltaJbar}) can be performed at NLL accuracy using the coupling defined in Eq.~(\ref{eq:rc-decoupling}) with its running at two loops:
\begin{equation}
	 \as^{(n)}(\kt^2)=\frac{\as^{(n)}(\mu_{n}^2)}{\ell}\left(1-\frac{\beta_1^{(n)}}{\beta_0^{(n)}}\as^{(n)}(\mu_{n}^2)\frac{\log{\left(\ell\right)}}{\ell}\right),
\end{equation}
where $\ell=1+\as^{(n)}(\mu_{n}^2)\beta^{(n)}_0\log{\frac{\kt^2}{\mu_{n}^2}}$, $\beta_0^{(n)}=\frac{11 \ca -2 n}{12 \pi}$, and $\beta_1^{(n)}=\frac{17\ca^2-5 \ca n-3 \cf n}{24 \pi^2}$. 

The result is expressed in terms of coefficient functions $C^{(i)}_\text{thr}$ that have support on regions $(i)$ that are identified by the hierarchy of the scales.
In particular, for the values of the centre-of-mass energies, quark masses, and moments considered in this study ($N\le20$), the charm ratio computed in this approximation is
\begin{equation}\label{eq:R_GMR}
    R_\text{thr}=
    \frac{C^{(1)}_\text{thr}(N,Q_A^2,\mu_A^2)\Theta_{A,1}+C^{(2)}_\text{thr}(N,Q_A^2,\mu_A^2)\Theta_{A,2}}{\sum_{i=1}^{4}C^{(i)}_\text{thr}(N,Q_B^2,\mu_B^2)\Theta_{B,i}}
\end{equation}
with
\begin{align}\label{eq:thetas}
\Theta_{A,1}&= \Theta\left(\frac{Q_A}{m_b}-\bar N \right),\quad \Theta_{A,2}= \Theta\left(\bar N-\frac{Q_A}{m_b}\right), \nonumber\\ 
    \Theta_{B,1}&= \Theta\left(\frac{Q_B}{m_b}-\bar N \right),\quad
\Theta_{B,2}= \Theta\left(\bar N-\frac{Q_B}{m_b} \right)  \Theta\left(\frac{Q_B^2}{m_b^2}-\bar N \right), \nonumber\\
    \Theta_{B,3}&= \Theta\left(\bar N-\frac{Q_B^2}{m_b^2} \right)  \Theta\left(\frac{Q_B}{m_c}-\bar N \right), \nonumber\\
    \Theta_{B,4}&=\Theta\left(\bar N-\frac{Q_B}{m_c} \right)   \Theta\left(\frac{Q_B^2}{m_c^2}-\bar N \right).
\end{align}
Details of the calculation can be found in~\cite{Ghira:2023bxr}, see also~\cite{Caletti:2023spr} for the extension to the multiple-threshold case.
The renormalisation scales are chosen such that 
$\mu_{3}\simeq m_c$, $\mu_{4}\simeq m_b$, $\mu_{5}\simeq Q_{A,B}$. Furthermore, since we are interested in the perturbative behaviour of the coefficient function, we fix $\muf=Q_{A,B}$. 
The result Eq.~(\ref{eq:R_GMR}) allows us to resum, in each region, logarithms of $\bar N$ and $\frac{m^2_{b,c}}{Q^2_{A,B}}$ to NLL accuracy.

Ideally, we would like to match the coefficient function computed in Eq.~(\ref{eq:GMR-start}) with the one that appears in the theoretical prediction for the charm ratio in Eq.~(\ref{eq:ratio_cntd}). However, while the latter is known to NNLL accuracy, the calculation performed in~\cite{Ghira:2023bxr} upon which Eq.~(\ref{eq:R_GMR}) only holds at NLL and its generalisation beyond this accuracy appears far from trivial. Nevertheless, we can note that the calculation of the charm ratio in the first region is the same as the one that appears in Eq.~(\ref{eq:ratio_cntd}), albeit at a lower accuracy. This suggests to define a $K$-factor through
\begin{equation}\label{eq:kfactor-def}
R_\text{thr}=\frac{C_\text{thr}^{(1)}(N,Q_A^2,\mu_A^2)}{C_\text{thr}^{(1)}(N,Q_B^2,\mu_B^2)} \times K_\text{thr}, \end{equation}
with
\begin{equation}\label{eq:kfactor}
K_\text{thr}=\frac{\Theta_{A,1}+r^{(2)}(N,Q_A^2,\mu_A^2)\Theta_{A,2}}{\sum_{i=1}^{4}r^{(i)}(N,Q_B^2,\mu_B^2)\Theta_{B,i}}, \; \; \text{and}
\; \;  r^{(i)}=\frac{C_\text{thr}^{(i)}}{C_\text{thr}^{(1)}}.
\end{equation}
We note that, for the kinematics that we are considering, $r^{(i)}>1$. Furthermore, the correction $r^{(2)}$ to the numerator of the $K$-factor is numerically rather small, in contrast to the ones in the denominator. Consequently, $K_\text{thr}<1$.
The $K$-factor's perturbative uncertainty is evaluated by varying the renormalisation scales $\mu_3$, $\mu_4$, and $\mu_5$ in the numerators of the ratios $r^{(i)}$, by a factor of two, up and down, about their natural scales $Q_A$ and $Q_B$
\footnote{Alternatively, we could have varied the scales in the denominators of $r^{(i)}$ too, keeping the variation correlated to avoid overestimating the uncertainties. However, with this procedure, the combination of the uncertainty with the NNLO+NNLL calculations would have been less straightforward.}.

\begin{figure}
	\centering
\includegraphics[width=0.49\textwidth]{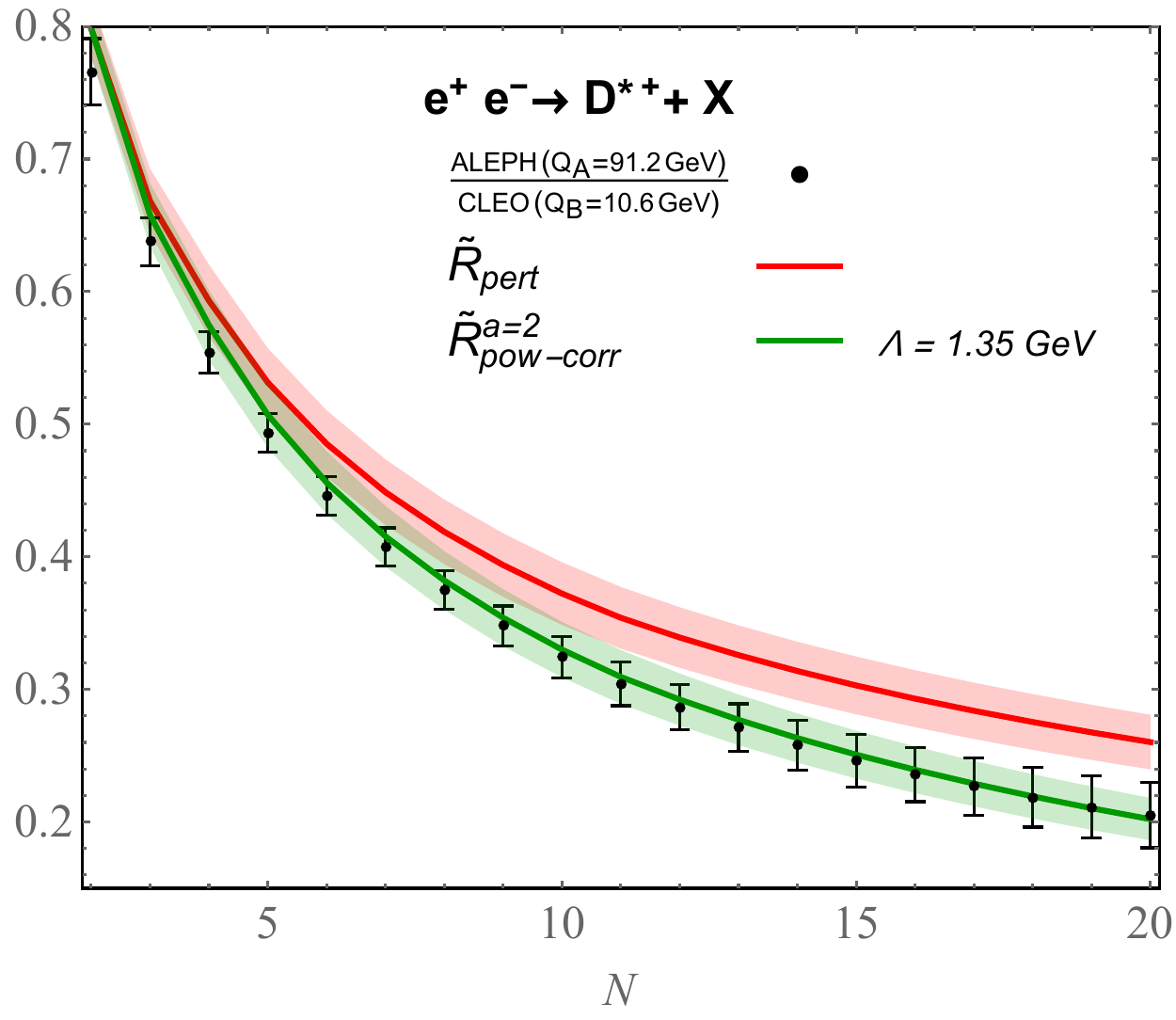}
\vspace{-0.5cm}
	\caption{Same as Fig.~\ref{fig:bcs}, but with the improved perturbative result $\widetilde{R}_\text{pert}$, Eq.~(\ref{eq:BCSxGMR}), in red, and $ \widetilde{R}_\text{pow-corr}^{a=2}$, Eq.~(\ref{eq:tildeR_nonpert}), with fitted $\Lambda=1.35$~GeV, in green.}
	\label{fig:bcsxgmr}
\end{figure}

We can use the $K$-factor to correct Eq.~(\ref{eq:ratio_cntd}) as follows
\begin{equation}\label{eq:BCSxGMR}
\widetilde{R}_\text{pert}=R_\text{pert} \times K_\text{thr}.
\end{equation}
Because $K_\text{thr}<1$, we expect this improved theoretical prediction for the charm ratio to move in the direction of the data.
The prediction obtained from Eq.~(\ref{eq:BCSxGMR}) is shown in Fig.~\ref{fig:bcsxgmr}. The uncertainty band is obtained by combining in quadrature the uncertainties from the NNLO+NNLL calculation, Eq.~(\ref{eq:ratio_cntd}) and from the $K$-factor, Eq.~(\ref{eq:kfactor}).
We note that the discrepancy between the improved perturbative prediction and the experimental data is noticeably reduced compared to what we have seen in Fig.~\ref{fig:bcs}. NLL corrections due to quark thresholds in the running of the coupling indeed account for more than half of the difference between theory and experiment.
This is almost entirely driven by the refined treatment of quark thresholds at the lower-energy experiment, CLEO, in this case. Indeed at higher scales, the $b$ and $c$ quarks behave as massless.
We expect the situation to further improve when yet-unknown NNLL effects are included. 

Meanwhile, it is instructive to repeat the study about non-perturbative power corrections with our improved perturbative predictions. Therefore, we define $\widetilde{R}_\text{pow-corr}^{a}$ using Eq.~(\ref{eq:R_nonpert}) but with $\widetilde{R}_\text{pert}$ instead of $R_\text{pert}$:
\begin{equation}\label{eq:tildeR_nonpert}
    \widetilde{R}_\text{pow-corr}^{a}=\widetilde{R}_\text{pert}\times \frac{1+\frac{\Lambda^a}{Q_A^a}\mathcal{C}(N)}{1+\frac{\Lambda^a}{Q_B^a}\mathcal{C}(N)}.
\end{equation}
Unsurprisingly, the fit to the experimental data now provides us with a reduced value for the non-perturbative scale $\Lambda=1.35$~GeV.
%, which is closer to the expected value of $\simeq 1$~GeV. 
%
Furthermore, the resulting $\widetilde{R}_\text{pert}$ appears to describe the data qualitatively better than $R_\text{pert}$. Because of the different theoretical uncertainties in the two fits, it is difficult to assess this quantitatively. However, a simple $\chi^2$ test using only the central values does confirm that the fit using the improved theory has better quality.

To summarise, we have considered the so-called charm ratio, which is constructed with Mellin moments of the $D^{*+}$ spectrum in $e^+e^-$ collisions, at different centre-of-mass energies. It has been argued in the literature that this ratio provides a direct test of time-like DGLAP evolution. However, the comparison between (N)NLO+(N)NLL theoretical predictions to ALEPH/CLEO data resulted in a long-standing discrepancy~\cite{Cacciari:2005uk,Bonino:2023icn}.
We have improved the all-order calculation by including in the resummed coefficient function heavy-quark mass effects recently derived at NLL in~\cite{Ghira:2023bxr}. Our new theoretical prediction is much closer to the experimental data, hence reducing the size of the non-perturbative power corrections needed to describe them well.
Given the numerical size of the effect, we believe that unknown NNLL corrections, in the decoupling scheme, could help further improve the agreement between theory and experiments. This motivates us to pursue this calculation.

Finally, we note that all-order threshold effects discussed here are also present in light-quark fragmentation. Although we expect their contribution to be diluted when singlet and non-singlet components are considered together, it would be interesting to investigate their impact on global fits~\cite{Borsa:2022vvp,Hernandez-Pinto:2017jyv,deFlorian:2014xna,Moffat:2021dji, Bertone:2017tyb,AbdulKhalek:2022laj}.

%\newpage
\paragraph{Acknowledgments.}
We thank Leonardo Bonino, Emanuele Nocera, and Giovanni Stagnitto for useful discussions and critical reading of the manuscript.
AG and SM would like to thank IPhT Saclay and LPTHE Paris for their hospitality at the start of this project.
This work is partly supported by the Italian Ministry of Research (MUR) under grant PRIN 2022SNA23K.

\bibliographystyle{jhep}      

\bibliography{charmratio}

\end{document}